\documentclass[preprint,tightenlines,showpacs,amsmath,amssymb]{revtex4}

\usepackage{epsfig,epsf,graphics,psfrag}
\usepackage{times}
\usepackage{float}
\usepackage{color}
\usepackage{amsfonts,amssymb,stmaryrd,latexsym,amsmath}
\usepackage{multirow}
\usepackage{array} 
\usepackage{hhline}
\usepackage{booktabs}
\usepackage{enumerate}
\usepackage{slashed}

\begin{document}

\bibliographystyle{unsrt}

\title{Understanding the nature of the heavy pentaquarks and searching for them in pion-induced reactions}

\author{Xiao-Hai Liu$^1$\footnote{liuxh@th.phys.titech.ac.jp} and Makoto Oka$^{1, 2}$\footnote{oka@th.phys.titech.ac.jp}}

\affiliation{ $^1$Department of Physics, H-27, Tokyo Institute of Technology, Meguro, Tokyo 152-8551, Japan}

\affiliation{$^2$ Advanced Science Research Center, JAEA, Tokai, Ibaraki 319-1195, Japan}

\date{\today}

\begin{abstract}
We investigate the reaction $\pi^- p \to \pi^- J/\psi p $ via the open-charm hadron rescattering diagrams. Due to the presence of the triangle singularity (TS) in the rescattering amplitudes, the TS peaks can simulate the pentaquark-like resonances arising in the $J/\psi p$ invariant mass distributions, which may bring ambiguities on our understanding of the nature of the exotic states. Searching for the heavy pentaquark in different processes may help us to clarify the ambiguities, because of the highly process-dependent characteristic of the TS mechanism.


\end{abstract}

\maketitle

\section{Introduction}
The LHCb collaborations recently reported the observations of two resonance-like structures in the invariant mass spectrum of $J/\psi p$ in the decay $\Lambda_b^0\to K^- J/\psi p$, which could be the long-searching-for pentaquark states~\cite{Aaij:2015tga}. The narrower pentaquark candidate $P_c(4450)$ has a mass of 4449.8$\pm$1.7$\pm$2.5 MeV and a width of 39$\pm$5$\pm$19 MeV, while the pentaquark candidate $P_c(4380)$ has a mass of 4380$\pm$8$\pm$29 MeV and a much broader width of 205$\pm$18$\pm$86 MeV \cite{Aaij:2015tga}.
These observations have intrigued a lot of studies regarding the underlying structures of the two exotic states. Many interpretations were proposed in the literature, such as the meson-baryon molecular states \cite{Wu:2010jy,Chen:2015loa,Roca:2015dva,Chen:2015moa,Karliner:2015ina,He:2015cea,Meissner:2015mza,Burns:2015dwa}, the pentaquark states composed of two diquarks and one antiquark (or one diquark and one triquark) \cite{Maiani:2015vwa,Li:2015gta,Anisovich:2015cia,Ghosh:2015ksa,Maiani:2015iaa,Cheng:2015cca,Wang:2015epa,Wang:2015ixb,Lebed:2015tna}, the kinematic threshold effects \cite{Liu:2015fea,Guo:2015umn,Mikhasenko:2015vca}, and so on. Some production modes to search for these states were also proposed, such as the photoproduction processes \cite{Wang:2015jsa,Kubarovsky:2015aaa,Karliner:2015voa}, or pion-induced reactions \cite{Garzon:2015zva,Wang:2015qia,Xie:2015zga,Lu:2015fva}.

When studying these heavy pentaquark candidates, usually one will confront two issues, i.e., what their underlying structures are and how to search for them in experiments. In Ref.~\cite{Liu:2015fea}, we pointed out that these resonance-like peaks may be resulted from some kinematic threshold effects, in particular the triangle singularity (TS) mechanism. Some similar studies can also be found in Refs.~\cite{Guo:2015umn,Mikhasenko:2015vca}.
The TS mechanism is a highly process-dependent mechanism, which is very different from other dynamic mechanisms. This may bring ambiguities on our understanding of the nature of those exotic states. We therefore need different kinds of processes to check this mechanism. The $\pi N$ collisions could be a promising reaction to search for the heavy pentaquark or the effects induced by the TS mechanism. An experiment to study the open-charm hadrons has been proposed at the forthcoming J-PARC high-momentum beam line \cite{Shirotori:2015eqa}, which may offer a good opportunity to study the heavy pentaquark.

The paper is organized as follows: In Section II, we will firstly give an introduction to the TS mechanism; The explicit formalism of our model, including the rescattering diagrams and the construction of the effective Lagrangians, will be described in Section II (A) and (B); The numerical results will be given in Section II (C).

\section{TS mechanism}
\begin{figure}[b]
	\centering
	\includegraphics[width=0.38\hsize]{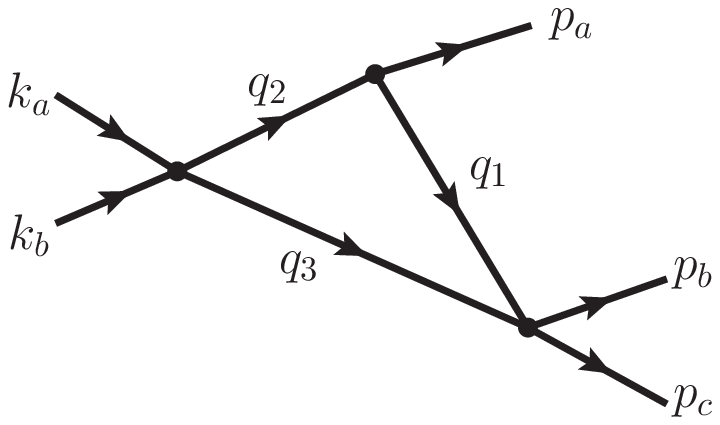}
	\caption{Triangle diagram under discussion. The internal mass which corresponds to the internal momentum $q_i$ is $m_i$ ($i$$=$1, 2, 3). The momentum symbols also represent the corresponding particles.}\label{triangle}
\end{figure}

 The possible manifestation of the kinematic singularities of the $S$-matrix elements was first noticed in 1960s and theoretical attempts were made to try to clarify the resonance-like structure, i.e. whether they are caused by the kinematic singularities or they are genuine resonance peaks~\cite{Peierls:1961zz,Goebel:1964zz,hwa:1963aa,landshoff:1962aa,Aitchison:1969tq,Aitchison:1964ak,coleman:1965aa,bronzan:1964aa,fronsdal:1964aa,norton:1964aa,schmid:1967aa}. Unfortunately, most of those proposed cases were lack of experimental support at that time. The TS mechanism was rediscovered by people in recent years and used to interpret some exotic phenomena, such as the largely isospin violations in $\eta(1405)$$\to$$3\pi$, the production of $a_1(1420)$, the production of some $XYZ$ particles and so on~\cite{Wu:2011yx,Ketzer:2015tqa,Wang:2013cya,Liu:2013vfa,Liu:2014spa,Szczepaniak:2015eza,Guo:2014iya,Liu:2015taa,Liu:2015cah}.
This mechanism is based on the study of the analytic properties of the $S$-matrix. For the triangle rescattering diagram displayed in Fig.~\ref{triangle}, in some special kinematic configurations, all of the three intermediate states can be on-shell simultaneously, which corresponds to the leading Landau singularity of the triangle diagram. This leading Landau singularity is usually called the TS. The physical picture concerning the TS is: The initial states $k_a$ and $k_b$ firstly scatter into particles $q_2$ and $q_3$, then the particle $q_1$ emitted from $q_2$ catches up with $q_3$, and finally $q_2$ and $q_3$ will rescatter into particles $p_b$ and $p_c$. This implies that the triangle rescattering diagram can be interpreted as a classical process in pace-time when the TS occurs, and the TS will be located on the physical boundary of the scattering amplitude~\cite{coleman:1965aa}. In Fig.~\ref{triangle}, we define the invariants $s_1=(k_a+k_b)^2$, $s_2=(p_b+p_c)^2$ and $s_3=p_a^2$. The locations of the TS can be determined by solving the Landau equations \cite{Landau:1959fi,Eden:1966,bonnevay:1961aa}. For instance,
if we fix the internal masses $m_i$, the external invariants $s_2$ and $s_3$, we can obtain the solutions for $s_1$, i.e.,
\begin{eqnarray}
s_1^{\pm}&=&(m_2+m_3)^2+\frac{1}{2m_1^2} {\LARGE[}(m_1^2+m_2^2-s_3)(s_2-m_1^2-m_3^2)-4m_1^2 m_2 m_3 \nonumber \\  &\pm& \lambda^{1/2}(s_2,m_1^2,m_3^2)\lambda^{1/2}(s_3,m_1^2,m_2^2){\LARGE ]},
\end{eqnarray}
with $\lambda(x,y,z)\equiv (x-y-z)^2-4yz$.
Likewise, by fixing $m_i$, $s_1$ and $s_3$ we can obtain the similar solutions for $s_2$, i.e.,
\begin{eqnarray}
s_2^{\pm}&=&(m_1+m_3)^2+\frac{1}{2m_2^2} {\LARGE[}(m_1^2+m_2^2-s_3)(s_1-m_2^2-m_3^2)-4m_2^2 m_1 m_3 \nonumber \\  &\pm& \lambda^{1/2}(s_1,m_2^2,m_3^2)\lambda^{1/2}(s_3,m_1^2,m_2^2){\LARGE ]}.
\end{eqnarray}
By means of the dispersion representation of the triangle diagram, we learn that within the physical boundary only the solution of $s_1^-$ or $s_2^-$ corresponds to the TS of the amplitude, and we call $s_1^-$ and $s_2^-$ as the anomalous thresholds \cite{Liu:2015taa,Eden:1966,bonnevay:1961aa}. For convenient, we define the normal threshold $s_{1N}$ ($s_{2N}$) and the critical value $s_{1C}$ ($s_{2C}$) for $s_1$ ($s_2$) as follows \cite{Liu:2015taa}, 
\begin{eqnarray}\label{s1Ns1C}
&& s_{1N}=(m_2+m_3)^2,\ s_{1C}=(m_2+m_3)^2 +\frac{m_3}{m_1}[(m_2-m_1)^2-s_3], \\
&& s_{2N}=(m_1+m_3)^2,\ s_{2C}=(m_1+m_3)^2 +\frac{m_3}{m_2}[(m_2-m_1)^2-s_3].
\end{eqnarray}
If we fix $s_3$ and the internal masses $m_{1,2,3}$, when $s_1$ increases from $s_{1N}$ to $s_{1C}$, the anomalous threshold $s_2^-$ will move from $s_{2C}$ to $s_{2N}$. Likewise, when $s_2$ increases from $s_{2N}$ to $s_{2C}$, $s_1^-$ will move from $s_{1C}$ to $s_{1N}$. This is the kinematic region where the TS can be present.
We can also use the discrepancies between the normal and anomalous thresholds to indicate the TS kinematic region. The maximum values of these discrepancies read
\begin{eqnarray} \label{deltas1s2}
\Delta_{s_1}^{\max}&=&\sqrt{s_{1C}} - \sqrt{s_{1N}}\approx
\frac{m_3}{2m_1(m_2+m_3)}[(m_2-m_1)^2-s_3], \nonumber \\
\Delta_{s_2}^{\max}&=&\sqrt{s_{2C}} - \sqrt{s_{2N}}\approx
\frac{m_3}{2m_2(m_1+m_3)}[(m_2-m_1)^2-s_3].
\end{eqnarray} 
If the above discrepancies are larger, usually it may be easier in experiments to observe some effects induced by the TS in the corresponding rescattering process.

\section{ $\pi^- p \to \pi^- J/\psi p $ via the triangle rescattering  diagrams}

\begin{figure}[t]
	\centering
	\includegraphics[width=0.9\hsize]{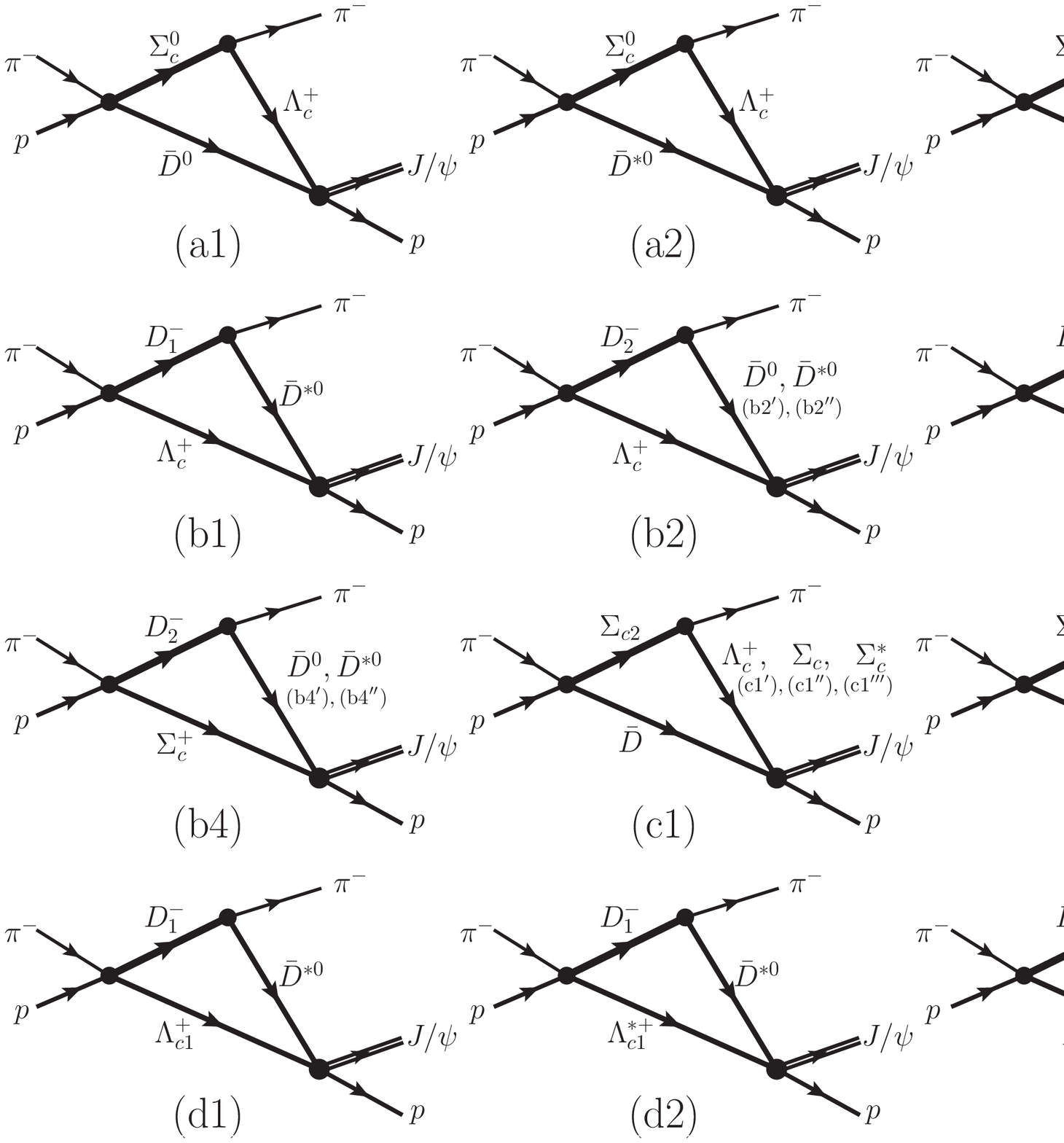}
	\caption{$\pi^- p \to \pi^- J/\psi p $ via the open-charm hadron rescattering  diagrams. The conventions of momentum and invariants are the same with those of Fig.~\ref{triangle}.}\label{pinscattering}
\end{figure}


\subsection{Possible rescattering processes recognizing the TS}
Motived by the observations of the heavy pentaquark candidates $P_c(4450)$ and $P_c(4380)$, we are going to apply the TS mechanism to the charmed meson and baryon systems. We firstly introduce some notations of the charmed hadrons in relevant with our discussions in Table~\ref{charmtable}.

\begin{table}\caption{Properties and notations of the charmed hadrons in relevant with our discussion. The experimental values are taken from the Particle Data Group (PDG)~\cite{Agashe:2014kda}.} 
	\begin{center}
		\begin{tabular}{c|c|c|c|c}
			\hline
			\hline
			Particle in PDG & Notation & $J^P$  & Mass [MeV] & Width [MeV] \\
			\hline
			$\Lambda_c^+$ & $\Lambda_c^+$ & $\frac{1}{2}^+$  & 2286.46$\pm$0.14 &  \\
			$\Lambda_c(2595)^+$ & $\Lambda_{c1}^+$ & $\frac{1}{2}^-$  & 2592.25$\pm$ 0.28  & 2.6$\pm$0.6 \\
			$\Lambda_c(2625)^+$ & $\Lambda_{c1}^{*+}$ & $\frac{3}{2}^-$  & 2628.11$\pm$0.19 & $<$0.97 \\
			$\Sigma_c(2455)^{++}$ & $\Sigma_c^{++}$ & $\frac{1}{2}^+$  & 2453.98$\pm$0.16 & 2.26$\pm$0.25 \\
			$\Sigma_c(2455)^{+}$ & $\Sigma_c^{+}$ & $\frac{1}{2}^+$  & 2452.9$\pm$0.4 & $<$4.6 \\	
			$\Sigma_c(2455)^0$ & $\Sigma_c^{0}$ & $\frac{1}{2}^+$  & 2453.74$\pm$0.16 & 2.16$\pm$0.26 \\
			$\Sigma_c(2520)^{++}$ & $\Sigma_c^{*++}$ & $\frac{3}{2}^+$  & 2517.9$\pm$0.6 & 14.9$\pm$1.5 \\
			$\Sigma_c(2520)^{+}$ & $\Sigma_c^{*+}$ & $\frac{3}{2}^+$  & 2517.5$\pm$2.3 & $<$17 \\
			$\Sigma_c(2520)^{0}$ & $\Sigma_c^{*0}$ & $\frac{3}{2}^+$  & 2518.8$\pm$0.6 & 14.5$\pm$1.5 \\	
			$\Sigma_c(2800)^{++}$ & $\Sigma_{c2}^{++}$ & $\frac{3}{2}^- ?$  & 2801$^{+4}_{-6}$ & 75$^{+22}_{-17}$ \\
			$\Sigma_c(2800)^{+}$ & $\Sigma_{c2}^{+}$ & $\frac{3}{2}^- ?$  & 2792$^{+14}_{-5}$ & 62$^{+60}_{-40}$ \\
			$\Sigma_c(2800)^{0}$ & $\Sigma_{c2}^{0}$ & $\frac{3}{2}^- ?$  & 2806$^{+5}_{-7}$ & 72$^{+22}_{-15}$ \\	
			\hline
			$D^0$ & $D^0$ & $0^-$  & 1864.84$\pm$0.07 &  \\	
			$D^{\pm}$ & $D^\pm$ & $0^-$  & 1869.61$\pm$0.10 &  \\	
			$D^*(2007)^{0}$ & $D^{*0}$ & $1^-$  & 2006.96$\pm$0.10 & $<$2.1  \\	
			$D^*(2010)^{\pm}$ & $D^{*\pm}$ & $1^-$  & 2010.26$\pm$0.07 & 0.0834$\pm$ 0.0018  \\	
			$D_1(2420)^{0}$ & $D_1^{0}$ & $1^+$  & 2421.4$\pm$0.6 & 27.4$\pm$2.5  \\
			$D_2(2460)^{0}$ & $D_2^{0}$ & $2^+$  & 2462.6$\pm$0.6 & 49.0$\pm$1.3  \\	
			$D_2(2460)^{\pm}$ & $D_2^{\pm}$ & $2^+$  & 2464.3$\pm$1.6 & 37$\pm$6  \\			
			\hline
		\end{tabular}
	\end{center}\label{charmtable}
\end{table} 

According to Eq.~(\ref{deltas1s2}), we learn the point that to enlarge the TS kinematic region, we need to enlarge the quantity $[(m_2-m_1)^2-s_3]$. Physically, this quantity corresponds to the phase space factor of the particle $q_2$ decaying into particles $p_a$ and $q_1$ in Fig.~\ref{triangle}. However, the larger phase space usually implies that the particle $q_2$ will be more unstable, and
if the intermediate state has a very large decay width, the TS will be removed far away from the physical boundary, which will weaken the influence of the TS over the physical amplitude \cite{Aitchison:1964ak,Wang:2013cya,Liu:2013vfa,Liu:2014spa}.
We therefore need a balance between the phase space and the stability of the intermediate state. In another word, to observe the resonance-like peaks induced by the TS, we will require that the phase space of $q_2$ decaying into $p_a$ and $q_1$ is relatively large, and at the same time $q_2$ should not be a very broad resonance. Taking into account the above requirements, some possible rescattering processes where the TS may be detected are illustrated in Fig.~\ref{pinscattering} for the pion-induced reaction $\pi^- p \to \pi^- J/\psi p $.
 
In Fig.~\ref{pinscattering}, the rescattering diagrams  $\pi^- p \to \bar{\mathcal{D}} Y_c \to \pi^- J/\psi p $ are divided into four categories according to the quantum numbers of the intermediate anti-charmed meson $\bar{\mathcal{D}}$ and charmed baryon $Y_c$, i.e.,
\begin{enumerate}
\item[I)] (a1)-(a3): Both $\bar{\mathcal{D}}$ and $Y_c$ are $S$-wave states,
\item[II)] (b1)-(b4): $\bar{\mathcal{D}}$ is a $P$-wave state and $Y_c$ is an  $S$-wave state,
\item[III)] (c1)-(c2): $\bar{\mathcal{D}}$ is an $S$-wave state and $Y_c$ is a  $P$-wave state,
\item[IV)] (d1)-(d3): Both $\bar{\mathcal{D}}$ and $Y_c$ are $P$-wave states.
\end{enumerate}
The kinematic regions where the TS can be present are listed in Table.~\ref{TSregion} for each of these diagrams.
\begin{table}\caption{The TS kinematic regions corresponding to the diagrams in Fig.~\ref{pinscattering}, in the unit of MeV.} 
	\begin{center}
		\begin{tabular}{c|c|c|c|c|c|c}
			\hline
			\hline
	Diagram \# & $ \sqrt{s_{1N}} $ & $ \sqrt{s_{1C}} $ & $ \Delta _{s_1}^{\max } $ & $ \sqrt{s_{2N}} $ & $ \sqrt{s_{2C}} $ & $ \Delta _{s_2}^{\max } $ \\
	\hline
	(a1) & 4317.7 & 4318.5 & 0.773 & 4151.4 & 4152.1 & 0.75 \\
	(a2) & 4459.9 & 4460.7 & 0.806 & 4293.5 & 4294.3 & 0.78 \\
	(a3) & 4524.5 & 4527.7 & 3.28 & 4293.5 & 4296.7 & 3.14 \\
	(b1) & 4708.0 & 4726.3 & 18.4 & 4293.5 & 4310.2 & 16.7 \\
	(b2$^\prime$) & 4750.9 & 4794.5 & 43.7 & 4151.4 & 4189.2 & 37.8 \\
	(b2$^{\prime\prime}$) & 4750.9 & 4773.5 & 22.7 & 4293.5 & 4314.0 & 20.4 \\
	(b3) & 4874.3 & 4893.4 & 19.1 & 4459.9 & 4477.1 & 17.3 \\
	(b4$^\prime$) & 4917.2 & 4962.4 & 45.2 & 4317.7 & 4356.7 & 39.0 \\
	(b4$^{\prime\prime}$) & 4917.2 & 4940.7 & 23.5 & 4459.9 & 4481.0 & 21.1 \\
	(c1$^\prime$) & 4665.8 & 4687.2 & 21.4 & 4151.4 & 4171.0 & 19.6 \\
	(c1$^{\prime\prime}$) & 4665.8 & 4674.1 & 8.28 & 4317.7 & 4325.6 & 7.83 \\
	(c1$^{\prime\prime\prime}$) & 4665.8 & 4670.7 & 4.83 & 4382.3 & 4387.0 & 4.62 \\
	(c2$^\prime$) & 4808.0 & 4830.3 & 22.3 & 4293.5 & 4313.9 & 20.4 \\
	(c2$^{\prime\prime}$) & 4808.0 & 4816.6 & 8.65 & 4459.9 & 4468.0 & 8.16 \\
	(c2$^{\prime\prime\prime}$) & 4808.0 & 4813.0 & 5.05 & 4524.5 & 4529.3 & 4.82 \\
	(d1) & 5013.7 & 5033.2 & 19.6 & 4599.2 & 4616.9 & 17.7 \\
	(d2) & 5049.5 & 5069.2 & 19.7 & 4635.1 & 4652.9 & 17.8 \\
	(d3$^\prime$) & 5092.4 & 5139.2 & 46.8 & 4493.0 & 4533.1 & 40.2 \\
	(d3$^{\prime\prime}$) & 5092.4 & 5116.7 & 24.3 & 4635.1 & 4656.8 & 21.8 \\	
			\hline
		\end{tabular}
	\end{center}\label{TSregion}
\end{table} 
From Table~\ref{TSregion}, we can see that the diagrams of Fig.~\ref{pinscattering} which involve $\bar{D}_1$, $\bar{D}_2$ or $\Sigma_{c2}$ in the loops have the relatively lager TS kinematic regions. This is because the phase spaces for $\bar{D}_1\to \bar{D}^*\pi$, $\bar{D}_2\to \bar{D}^{(*)}\pi$ and $\Sigma_{c2}\to \Lambda_c\pi / \Sigma_c^{(*)}\pi $ are sizable, as discussed in the above section.

\subsection{Formalism}
To quantitatively estimate the rescattering amplitudes corresponding to Fig.~\ref{pinscattering}, we will build our model in the framework of heavy hadron chiral
perturbation theory (HHChPT)~\cite{Casalbuoni:1992dx,Casalbuoni:1996pg,Colangelo:2003sa,Cho:1994vg,Pirjol:1997nh,Cheng:2006dk,Cho:1992gg,Colangelo:2012xi}. In HHChPT, to encode the heavy quark spin symmetry (HQSS), the charmed meson doublets $(D,\ D^*)$ and $(D_1,\ D_2)$ are collected into the following superfields
\begin{eqnarray}
&& H_{1a}  = \frac{1+\slashed{v}}{2}[D_{a\mu}^*\gamma^\mu-D_a\gamma_5] , \\
&& H_{2a}  = [\bar{D}_{a\mu}^*\gamma^\mu+\bar{D}_a\gamma_5]\frac{1-\slashed{v}}{2} , \\
&& T_{1a}^\mu = \frac{1+\slashed{v}}{2} \bigg\{ D^{\mu\nu}_{2a} \gamma_\nu -\sqrt{3 \over 2}D_{1a\nu} \nonumber \\
&&\times \gamma_5 \left[
g^{\mu \nu}-{1 \over 3} \gamma^\nu (\gamma^\mu-v^\mu) \right]
\bigg\} \ , \\
&& T_{2a}^\mu = \bigg\{ \bar{D}^{\mu\nu}_{2a} \gamma_\nu +\sqrt{3 \over 2}\bar{D}_{1a\nu} \nonumber \\
&& \times  \gamma_5 \left[
g^{\mu \nu}-{1 \over 3} \gamma^\nu (\gamma^\mu-v^\mu) \right]
\bigg\}\frac{1-\slashed{v}}{2} \ , \\
&& \bar{H}_{1a,2a} = \gamma^0 H_{1a,2a}^{\dag} \gamma^0,\ \bar{T}_{1a,2a} = \gamma^0 T_{1a,2a}^{\dag} \gamma^0,
\end{eqnarray}
where $H_{2a}$ ($T_{2a}$) is the charge conjugate field of $H_{1a}$ ($T_{1a}$), and $a$ is the light flavor index. The charmed baryons $\Lambda_c$, $\Lambda_{c1}$, $\Lambda_{c1}^*$, $\Sigma_c$, $\Sigma_c^*$ and $\Sigma_{c2}$ are collected in the following superfileds
\begin{eqnarray}
&& \mathcal{T}_i=\frac{1}{2} \epsilon_{ijk} \frac{1+\slashed{v}}{2} (B_{\bar{3}})_{jk}, \\
&& \mathcal{R}_{\mu}=\frac{1}{\sqrt{3}} (\gamma_\mu +v_\mu)\gamma_5  \frac{1+\slashed{v}}{2} \Lambda_{c1}^{+} + \frac{1+\slashed{v}}{2} \Lambda_{c1\mu}^{*+}, \\
&& \mathcal{S}_{\mu}^{ij}=\frac{1+\slashed{v}}{2} B_{6\mu}^{*ij} + \frac{1}{\sqrt{3}} (\gamma_\mu +v_\mu)\gamma_5  \frac{1+\slashed{v}}{2} B_{6}^{ij}, \\
&& \mathcal{X}_{\mu\nu}^{ij}= \frac{1}{\sqrt{10}} \big[  (\gamma_\mu +v_\mu)\gamma_5 g_\nu^\alpha + (\gamma_\nu +v_\nu)\gamma_5 g_\mu^\alpha
\big] X_{\alpha}^{ij},
\end{eqnarray}
with the matrices
\begin{eqnarray}
(B_{\bar{3}})_{ij}= \left(
\begin{array}{ccc}
0 & \Lambda_c^{+} &\Xi_c^{+} \\
-\Lambda_c^{+} & 0  & \Xi_c^{0} \\
-\Xi_c^{+} & -\Xi_c^{0} & 0 \\
\end{array}
\right)_{ij},
\end{eqnarray}

\begin{eqnarray}
(B_{6})_{ij}= \left(
\begin{array}{ccc}
\Sigma_c^{++} & \frac{1}{\sqrt{2}} \Sigma_c^{+} & \frac{1}{\sqrt{2}}\Xi_c^{\prime +} \\
\frac{1}{\sqrt{2}} \Sigma_c^{+} & \Sigma_c^{0} & \frac{1}{\sqrt{2}}\Xi_c^{\prime 0} \\
\frac{1}{\sqrt{2}}\Xi_c^{\prime +} & \frac{1}{\sqrt{2}}\Xi_c^{\prime 0} & \Omega_c^0 \\
\end{array}
\right)_{ij},
\end{eqnarray}
\begin{eqnarray}
(X)_{ij}= \left(
\begin{array}{ccc}
\Sigma_{c2}^{++} & \frac{1}{\sqrt{2}} \Sigma_{c2}^{+} & \frac{1}{\sqrt{2}}\Xi_{c2}^{\prime +} \\
\frac{1}{\sqrt{2}} \Sigma_{c2}^{+} & \Sigma_{c2}^{0} & \frac{1}{\sqrt{2}}\Xi_{c2}^{\prime 0} \\
\frac{1}{\sqrt{2}}\Xi_{c2}^{\prime +} & \frac{1}{\sqrt{2}}\Xi_{c2}^{\prime 0} & \Omega_{c2}^0 \\
\end{array}
\right)_{ij}.
\end{eqnarray}

For the charmed hadron decaying into one pion and another charmed hadron, we have the following effective Lagrangians which respect the HQSS:
\begin{eqnarray}
 \mathcal{L}_{\mbox{meson}} &=&  i{h^\prime \over \Lambda_\chi} \mbox{Tr} \big[ {\bar H}_{2a}
T^\mu_{2b} \gamma^\nu\gamma_5 ( D_\mu \mathcal{A}_\nu + D_\nu
\mathcal{A}_\mu)_{ba} \big] +\ \mbox{h.c.}, \label{lagmeson}\\
\mathcal{L}_{\mbox{baryon}} &=& -\sqrt{3} g_2 \mbox{Tr} \big[ \bar{B}_{\bar{3}} \mathcal{A}^\mu \mathcal{S}_\mu + \bar{\mathcal{S}}_\mu \mathcal{A}^\mu B_{\bar{3}} \big] \nonumber \\
&+& i h_{10} \epsilon_{ijk}\bar{\mathcal{T}}_i (D_\mu \mathcal{A}_\nu + D_\nu
\mathcal{A}_\mu)_{jl} \mathcal{X}_{kl}^{\mu\nu}  \nonumber \\
&+& h_{11} \epsilon_{\mu\nu\sigma\lambda}  v^\lambda  \mbox{Tr} \big[ \bar{\mathcal{S}}^\mu  (D^\nu \mathcal{A}_\alpha + D_\alpha
\mathcal{A}^\nu) \mathcal{X}^{\alpha\sigma} \big], \label{lagbaryon}
\end{eqnarray}
where $\mathcal{A}^\mu$ is the chiral axial vector containing the
Goldstone bosons 
\begin{eqnarray}
\mathcal{A}_\mu &=& \frac{i}{2} \left( \xi^\dag\partial_\mu\xi  - \xi\partial_\mu\xi^\dag  \right)\ ,
\end{eqnarray}
with
\begin{eqnarray}
\xi = e^{i\mathcal{M}/f_\pi}\ ,\ 
\mathcal{M}= \left(
\begin{array}{ccc}
\frac{1}{\sqrt{2}}\pi^0+ \frac{1}{\sqrt{6}}\eta & \pi^+ & K^{+} \\
\pi^- & -\frac{1}{\sqrt{2}}\pi^0+ \frac{1}{\sqrt{6}}\eta  & K^{0} \\
K^{-} & \bar{K}^{0} & -\sqrt{\frac{2}{3}}\eta \\
\end{array}
\right),
\end{eqnarray}
and the covariant derivative is defied as $D_\mu$$=$$\partial_\mu +\mathcal{V}_\mu$, with the chiral vector
\begin{eqnarray}
\mathcal{V}_\mu = \frac{1}{2} \left( \xi^\dag\partial_\mu\xi  + \xi\partial_\mu\xi^\dag  \right).
\end{eqnarray}
The reader is also referred to Refs.~\cite{Cho:1994vg,Pirjol:1997nh,Cheng:2006dk,Cho:1992gg,Colangelo:2012xi} for more details concerning these effective Lagrangians. The coupling constants in the above interactions can be determined by the corresponding decay widths measured in experiments. We adopt the averaged values estimated in Refs.~\cite{Cheng:2006dk,Colangelo:2012xi,Cheng:2015naa},
\begin{eqnarray}
&& h^\prime = 0.43, \\
&& g_2 = 0.565, \\
&& |h_{10}| = 0.85.
\end{eqnarray}
By means of the quark model, the couplings $h_{10}$ and $h_{11}$ will satisfy the relation $|h_{11}|$$=$$\sqrt{2}|h_{10}|$ \cite{Pirjol:1997nh}. The chiral symmetry breaking scale is set to $\Lambda_\chi$=1 GeV, and the pion decay constant is taken as $f_\pi$= 132 MeV. From Eqs.~(\ref{lagmeson}) and (\ref{lagbaryon}), one may notice that the dominant two-body decay modes $\bar{D}_1\to \bar{D}^*\pi$, $\bar{D}_2\to \bar{D}^{(*)}\pi$ and $\Sigma_{c2}\to \Lambda_c\pi / \Sigma_c^{(*)}\pi $ are actually $D$-wave decays due to the HQSS and the angular momentum conservation, which makes the decaying particles become relatively stable, although the phase spaces of these decay modes are large enough. This is one important advantage to be able to observe the effects resulted by the TS mechanism.


The effective Lagrangian in relevant with the vertex $\bar{\mathcal{D}}Y_c \to J/\psi p$ takes the form
\begin{eqnarray}
\mathcal{L}_{\mbox{ct}}= g_{\Lambda_c} \bar{N} H_2 \bar{J} \mathcal{T}_3 + g_{\Sigma_c} \bar{N}\gamma_\mu\gamma_5 H_2 \bar{J} \mathcal{S}^\mu + ig_{\Lambda_{c1}} \partial_\mu \bar{N} H_2 \bar{J} \mathcal{R}^\mu, \label{contactJ}
\end{eqnarray} 
where $\bar{N}$ is the isospin doublet $(\bar{p},\ \bar{n})$, and $J$ indicates the $S$-wave charmonia
\begin{eqnarray}
&& J = \frac{1+\slashed{v}}{2} [\psi(nS)^\mu \gamma_\mu-\eta_c(nS)\gamma_5]
\frac{1-\slashed{v}}{2}\ , \nonumber \\
&& \bar{J}=\gamma^0 J \gamma^0.
\end{eqnarray}
For these short range interactions, from the dimensional analysis, it is expected that the couplings $g_{\Lambda_c}$ ($g_{\Sigma_c}$) and $g_{\Lambda_{c1}}$  are of the order of magnitude of $m_D^{-2}$ and $m_D^{-3}$ respectively, where $m_D$ is the mass of the $D$ meson. Notice that in HHChPT, the heavy filed $H_2$ ($J$) in Eq.~(\ref{contactJ}) will contain a factor $\sqrt{M_{H_2}}$ ($\sqrt{M_J}$) for normalization. In addition, we will estimate the scattering amplitudes in the static limit , which means the four velocity is set to $v=(1, 0, 0, 0)$.

For the vertex $\pi^- p \to \bar{\mathcal{D}}Y_c$ in Fig.~\ref{pinscattering}, the effective interactions involving the fewest derivatives are constructed as follows:
\begin{eqnarray}
\mathcal{L}_{\mbox{a1}} &=& \frac{g_{a1}}{2m_D}\  \mathbb{D}\  \mathbf{\tau}\cdot\mathbf{\pi} \  \mathbf{\tau}\cdot\mathbf{\bar{\Sigma}_c} \ N, \\
\mathcal{L}_{\mbox{a2}} &=& \frac{g_{a2}}{2m_D}\  \mathbb{D}^{*}_{\mu}\  \mathbf{\tau}\cdot\mathbf{\pi} \  \mathbf{\tau}\cdot\mathbf{\bar{\Sigma}_c} \gamma_5 \gamma^\mu N, \\
\mathcal{L}_{\mbox{a3}} &=& \frac{g_{a3}}{2m_D}\  \mathbb{D}^{*}_{\mu}\  \mathbf{\tau}\cdot\mathbf{\pi} \  \mathbf{\tau}\cdot\mathbf{\bar{\Sigma}_c^{*\mu}} \  N, \\
\mathcal{L}_{\mbox{b1}} &=& \frac{i g_{b1}}{\sqrt{2}m_D^2}\   \mathbb{D}_1^{\mu} \  \mathbf{\tau}\cdot \partial_\mu \mathbf{\pi} \ \bar{\Lambda}_c \   N ,\\
\mathcal{L}_{\mbox{b2}} &=& \frac{i g_{b2}}{\sqrt{2}m_D^2}\  \mathbb{D}^{\mu\nu}_{2}  \   \mathbf{\tau}\cdot \partial_\mu \mathbf{\pi} \ \bar{\Lambda}_c \gamma_5 \gamma_\nu N ,\\
\mathcal{L}_{\mbox{b3}} &=& \frac{i g_{b3}}{2m_D^2}\   \mathbb{D}_1^{\mu} \  \mathbf{\tau}\cdot \partial_\mu \mathbf{\pi} \ \mathbf{\tau}\cdot\mathbf{\bar{\Sigma}_c} \   N ,\\
\mathcal{L}_{\mbox{b4}} &=& \frac{i g_{b4}}{2m_D^2}\  \mathbb{D}^{\mu\nu}_{2}  \   \mathbf{\tau}\cdot \partial_\mu \mathbf{\pi} \ \mathbf{\tau}\cdot\mathbf{\bar{\Sigma}_c} \gamma_5 \gamma_\nu N ,\\
\mathcal{L}_{\mbox{c1}} &=& \frac{i g_{c1}}{2m_D^2}\  \mathbb{D}\  \mathbf{\tau}\cdot \partial_\mu\mathbf{\pi} \  \mathbf{\tau}\cdot\mathbf{\bar{\Sigma}_{c2}^{\mu}} \  N ,\\
\mathcal{L}_{\mbox{c2}} &=& \frac{i g_{c2}}{2m_D^2}\  \mathbb{D}^{*}_\mu\  \mathbf{\tau}\cdot \partial_\nu\mathbf{\pi} \  \mathbf{\tau}\cdot\mathbf{\bar{\Sigma}_{c2}^{\mu}} \gamma_5 \gamma^\nu N ,\\
\mathcal{L}_{\mbox{d1}} &=& \frac{g_{d1}}{\sqrt{2}m_D}\   \mathbb{D}_1^{\mu} \  \mathbf{\tau}\cdot \mathbf{\pi} \ \bar{\Lambda}_{c1} \gamma_5 \gamma_\mu   N ,\\
\mathcal{L}_{\mbox{d2}} &=& \frac{g_{d2}}{\sqrt{2}m_D}\   \mathbb{D}_1^{\mu} \  \mathbf{\tau}\cdot \mathbf{\pi} \ \bar{\Lambda}_{c1\mu}^{*} \   N ,\\
\mathcal{L}_{\mbox{d3}} &=& \frac{g_{d3}}{\sqrt{2}m_D}\  \mathbb{D}^{\mu\nu}_{2}  \   \mathbf{\tau}\cdot \mathbf{\pi} \ \bar{\Lambda}_{c1\mu}^{*} \gamma_5 \gamma_\nu N,
\end{eqnarray}
where $\mathbf{\tau}$, $\mathbb{D}$ and $N$ represent the usual Pauli matrix, isospin doublets $(D^0,\ D^+)$ and $(p,\ n)^T$, respectively. To our knowledge there is little information about the open-charm hadron near threshold production in $\pi N$ collisions, both in experiments and theories. The only available data come from the BNL more than thirty years ago, where the upper limit for each of the cross sections $\sigma(\pi^- p \to D^{*-}\Lambda_c^+)$ and $\sigma(\pi^- p \to D^{*-}\Sigma_c^+)$ at 13 GeV pion-beam energy is about 7 nb \cite{Christenson:1985ms}. In Refs.~\cite{Kim:2015ita,Kim:2014qha}, the authors estimated that the cross sections for the open-charm hadron near threshold production are at the order of magnitude of 1 nb, by means of an effective Lagrangian method and a Regge approach. A similar production rate was predicted within the generalized parton picture in Ref.~\cite{Kofler:2014yka}. The charmed baryon production cross section of 1 nb was also assumed for the J-PARC experimental design \cite{Shirotori:2015eqa}. To determine the coupling constants $g_{a1}$--$g_{d3}$, we therefore assume that each of the cross sections 
$\sigma(\pi^- p \to \bar{\mathcal{D}} Y_c)$ is 1 nb at 20 GeV pion-beam energy. According to this assumption, the estimated coupling constants are listed in Table~\ref{coupga1}.


\begin{table}\caption{Coupling constants $g_{a1} - g_{d3}$ adopted in our calculations, in the unit of 1.} 
	\begin{center}
		\begin{tabular}{|c|c|c|c|c|c|}
			\hline
			\hline
			 $g_{a1}$ & $g_{a2}$ & $g_{a3}$ & $g_{b1}$ & $g_{b2}$  & $g_{b3}$\\
		\hline	 		
		      0.035 & 0.020 & 0.025 & 0.024 & 0.018  & 0.023\\
		 \hline
		      $g_{b4}$ & $g_{c1}$ & $g_{c2}$ & $g_{d1}$ & $g_{d2}$  & $g_{d3}$\\
		      \hline
		      0.017 & 0.027 & 0.009 & 0.024 & 0.026  & 0.029\\		
			\hline
		\end{tabular}
	\end{center}\label{coupga1}
\end{table}

We should mention that the above estimations on the coupling constants of the two contact interactions $\pi^- p \to \bar{\mathcal{D}} Y_c$ and $\bar{\mathcal{D}} Y_c \to J/\psi p$ will have larger uncertainties, and we only expect that those crude estimations will make sense in the order of magnitude. Nevertheless, when the TS being present in the rescattering amplitude, the line-shape behavior of the corresponding invariant mass spectrum will mainly depend on the kinematics, which is model-independent.

Because there may be unstable particles appearing in the triangle rescattering diagrams, in order to account for the width effects of the intermediate states, we will adopt the Breit-Wigner type propagators when calculating the triangle loop integrals \cite{Aitchison:1964ak}. For instance, for the spin-0 charmed meson $\mathcal{D}$ and spin-$\frac{1}{2}$ charmed baryon $Y_c$, the propagators read
\begin{eqnarray}
G^{(0)}_\mathcal{D}=\frac{i}{q_\mathcal{D}^2-m_\mathcal{D}^2+i\ m_\mathcal{D} \Gamma_\mathcal{D}},
\end{eqnarray}  
and
\begin{eqnarray}
G^{(\frac{1}{2})}_{Y_c}=\frac{i\ m_{Y_c}(1+\slashed{v})}{q_{Y_c}^2-m_{Y_c}^2+i\ m_{Y_c} \Gamma_{Y_c}},
\end{eqnarray} 
respectively. The propagators of other higher-spin charmed hadrons will take the similar formalisms with different spin projection operators. The Rarita-Schwinger spin wave functions for particles of arbitray spin will be used in calculations \cite{Rarita:1941mf,Behrends:1957,Chung:1993da,Zou:2002yy}. The Breit-Wigner type propagator will remove the TS from the physical boundary by a small distance, if the corresponding decay width $\Gamma$ is smaller \cite{Aitchison:1964ak,schmid:1967aa}.  

\subsection{Numerical results}
\begin{figure}[tb]
	\centering
	\includegraphics[width=1.\hsize]{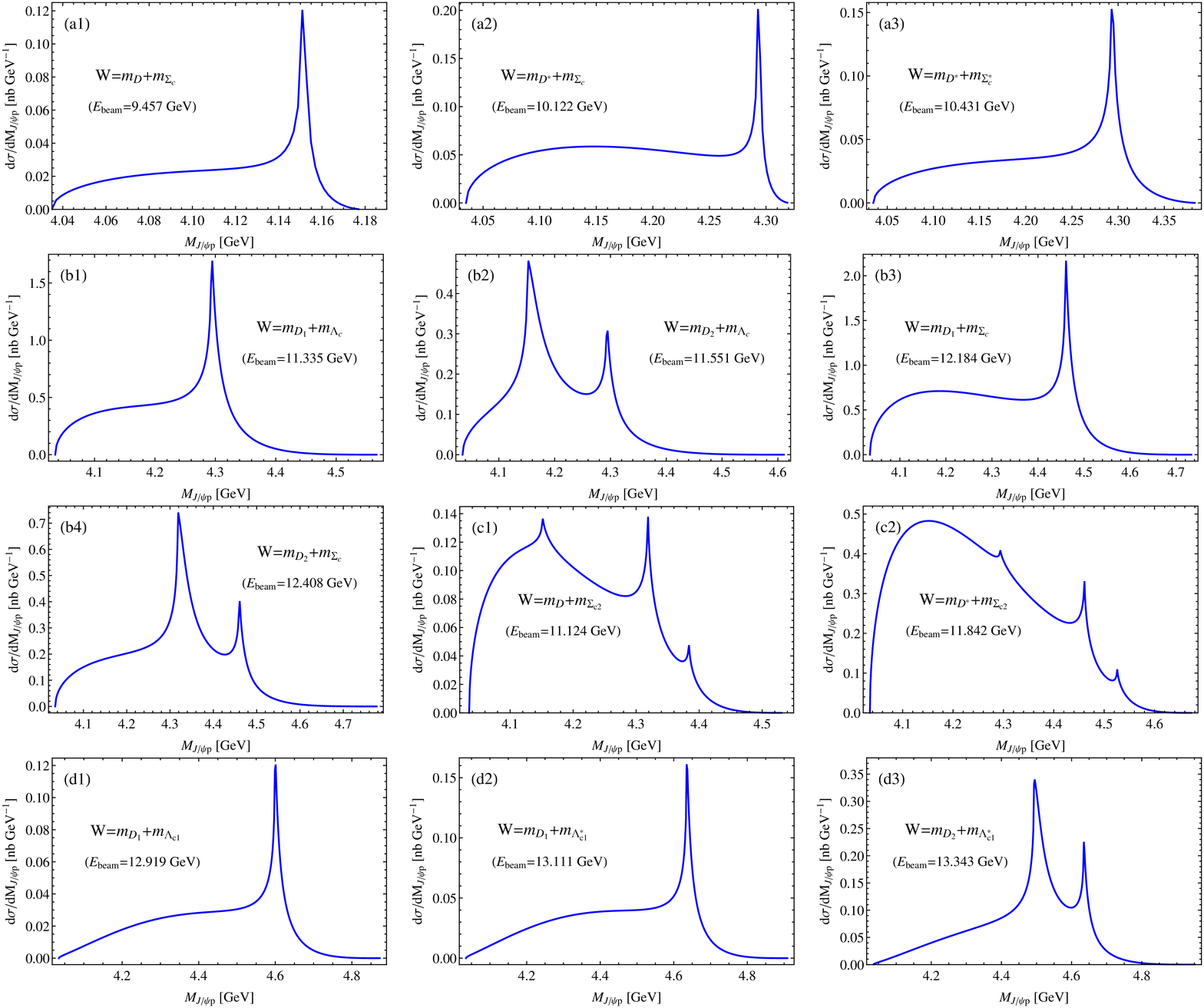}
	\caption{Invariant mass distributions of $J/\psi p$ in $\pi^- p \to \pi^- J/\psi p $ via the open-charm hadron rescattering diagrams in Fig.~\ref{pinscattering}. $W$ and $E_{\rm beam}$ represent the CM energy $\sqrt{s_1}$ and the corresponding pion-beam energy in the laboratory frame, respectively.}\label{invmdist}
\end{figure}

To observe the effects induced by the TS mechanism, we are particularly interested in the energy regions where the center-of-mass (CM) energies of $\pi^- p$ are located close to the $\bar{\mathcal{D}} Y_c$ thresholds for the reaction  $\pi^- p \to \pi^- J/\psi p $.  The numerical results for the invariant mass distributions of $J/\psi p$ in the process  $\pi^- p \to \pi^- J/\psi p $ via different triangle rescattering diagrams are displayed in Fig.~\ref{invmdist}. For each of the diagrams in Fig.~\ref{pinscattering}, the differential cross section is calculated at the corresponding $\bar{\mathcal{D}} Y_c$ threshold, and we did not take into account the interference terms between different diagrams. For instance, the $J/\psi p$ distribution displayed in Fig.~\ref{invmdist} (a1) is calculated at $\sqrt{s_1}$=$m_{D^0}+m_{\Sigma_c^0}$, and only the contribution from the diagram of Fig.~\ref{pinscattering} (a1) is taken into account. This is because for most of the diagrams in Fig.~\ref{pinscattering}, the kinematic regions of the TS in $s_1$ are well separated, which can be seen from Table~\ref{TSregion}. If we fix $\sqrt{s_1}$ at one specific $\bar{\mathcal{D}} Y_c$ threshold, usually the TS in $s_2$ can only be present in one of the diagrams, the contributions from other diagrams can then be taken as the background. Besides, the relative coupling strength and phases among those diagrams are actually not well determined in our model. However, for the diagrams in Fig.~\ref{pinscattering} of which the pertinent couplings are constrained by the HQSS, such as diagrams (b2$^\prime$) and (b2$^{\prime\prime}$), (b4$^\prime$) and (b4$^{\prime\prime}$), (c1$^{\prime\prime}$) and (c1$^{\prime\prime\prime}$), (c2$^{\prime\prime}$) and (c2$^{\prime\prime\prime}$), or (d3$^\prime$) and (d3$^{\prime\prime}$),  the interference terms have been included in calculations.

In Fig.~\ref{invmdist}, one may notice that many resonance-like peaks arise in the $J/\psi p$ invariant mass distributions, and these peaks stay around the $\bar{D}^{(*)}\Lambda_c$, $\bar{D}^{(*)}\Sigma_c$, $\bar{D}^{(*)}\Sigma_c^*$, $\bar{D}^{*}\Lambda_{c1}$ or $\bar{D}^{(*)}\Lambda_{c1}^*$ threshold. Besides, due to multiple decay modes of the intermediate state (particle $q_2$ in Fig.~\ref{triangle}), one may observe two or three peaks at the same CM energy of $\pi^- p$ collisions, such as in Figs.~\ref{invmdist} (b2), (b4), (c1), (c2) and (d3).
However, we did not introduce any genuine ``pentaquarks'' in the current model. 
These peaks are just induced by the TSs of the rescattering amplitudes, which implies a non-resonance explanation of the pentaquark candidate.
In addition, because of the proximity of these TS peaks to the meson-baryon thresholds, they may mix with the meson-baryon molecular states, which are supposed to be genuine resonances, if those molecular states truely exist.

The line shapes of the invariant mass spectrum in Fig.~\ref{invmdist} mainly depend on the kinematic configurations of the corresponding loop integrals, which will not be affected too much by the explicit coupling formalisms of the rescattering diagrams. However, if the backgrounds are extremely large compared with the contributions from the rescattering diagrams, the resonance-like peaks induced by the TS in Fig.~\ref{invmdist} may be absent. Here we use the word ``background'' to indicate the contributions from the processes without the $\bar{\mathcal{D}}Y_c$ rescatterings for the reaction $\pi^- p \to \pi^- J/\psi p $.

The production cross sections of the reaction $\pi^- p \to \pi^- J/\psi p $ via the rescatterings at different $\bar{\mathcal{D}} Y_c$ thresholds are listed in Table~\ref{totxec}. Those estimated cross sections are of the order of magnitude of $10^{-3}$ to $10^{-1}$ nb, which are not very large. One of the reasons is that we assume relatively lower production rates of the open-charm hadrons in $\pi N$ collisions. But taking into the high luminosity of the forthcoming J-PARC experiments, some effects induced by the TS mechanism may still be observed.

\begin{table}\caption{Total cross section of $\pi^- p \to \pi^- J/\psi p $ via the rescattering diagram in Fig.~\ref{pinscattering} at the corresponding $\bar{\mathcal{D}} Y_c$ threshold. }
	\begin{center}
\begin{tabular}{c|c}
	\text{Diagram \#} & \text{Cross Section [nb]} \\
	\hline
	\hline
	(a1) & 0.003 \\
	(a2) & 0.015 \\
	(a3) & 0.011 \\
	(b1) & 0.153 \\
	(b2) & 0.056 \\
	(b3) & 0.330 \\
	(b4) & 0.113 \\
	(c1) & 0.032 \\
	(c2) & 0.166 \\
	(d1) & 0.016 \\
	(d2) & 0.024 \\
	(d3) & 0.056 \\
	\hline
\end{tabular}
    \end{center}\label{totxec}
\end{table}


\section{Summary}
In this paper, we investigated the reactions $\pi^- p \to \pi^- J/\psi p $ via the open-charm hadron rescatterings. When the CM energies being taken around the $\bar{\mathcal{D}} Y_c$ thresholds, the TS may be present close to the physical boundary of the rescattering amplitudes. The TS peaks can then simulate the genuine resonances, which implies the possibility that some of the resonance-like peaks observed in experiments are resulted by the kinematic singularities.  
Being different from the genuine resonances, the TS peaks are rather sensitive to the kinematic configurations of the pertinent production processes. If the kinematic conditions of the TS are not satisfied, there will be no peaks appearing in the physical amplitudes. However,  one would expect that the genuine state should also appear in the processes where the kinematic conditions of the TS are not fulfilled. The presence of the TS means that a combined study of the TS mechanism and other dynamic processes are necessary. This should be crucial for our better understanding of those threshold enhancements. The forthcoming J-PARC pion-induced experiment may offer us a good opportunity to check different kinematic or dynamic mechanisms and clarify the ambiguities, with its high luminosity.

\subsection*{Acknowledgments}
Helpful discussions with A. Hosaka and Q. Zhao are gratefully acknowledged. X. H. Liu also thanks H. Noumi for the introduction of charmed baryon experiments in J-PARC. 
This work was supported by the Japan Society for the Promotion of Science under Contract No. P14324, and the JSPS KAKENHI (Grant No. 25247036).

\end{document}